\documentclass[runningheads]{llncs}

\usepackage[T1]{fontenc}
\usepackage{newtxtext,newtxmath}
\usepackage{etoolbox}
\AtBeginEnvironment{abstract}{\fontsize{9pt}{11pt}\selectfont}

\usepackage[T1]{fontenc}
\usepackage{caption}
\usepackage{subcaption}
\usepackage{amsmath,amsfonts}
\usepackage{algorithm}
\usepackage{algpseudocode}
\usepackage{array}
\usepackage[caption=false,font=normalsize,labelfont=sf,textfont=sf]{subfig}
\usepackage{textcomp}
\usepackage{stfloats}
\usepackage{url}
\usepackage{verbatim}
\usepackage{graphicx}

\usepackage{amssymb}
\usepackage{multirow}
\usepackage[table]{xcolor}
\definecolor{mycolor}{HTML}{FFFC9E}
\newcolumntype{a}{>{\columncolor{mycolor}}c}
\usepackage{cite}
\definecolor{mycolor2}{HTML}{9AFF99}
\newcolumntype{b}{>{\columncolor{mycolor2}}c}
\usepackage{xcolor}

\newcommand{\LOOPPE}{$\mathcal{LOOP-PE}$}

\usepackage{graphicx}
\begin{document}
\title{Towards Reliable Neural Optimizers: Permutation-Equivariant Neural Approximation in Dynamic Data-Driven Applications Systems}
%
%
\author{Meiyi Li\inst{1}\orcidID{0000-0002-0178-7883} \and
Javad Mohammadi\inst{1}\orcidID{0000-0003-0425-5302} }
\authorrunning{Meiyi Li,  Javad Mohammadi.}
%
\institute{The University of Texas at Austin,  TX , USA \\
\email{\{meiyil,javadm\}@utexas.edu}}
\maketitle              

\begin{abstract}
Dynamic Data Driven Applications Systems (DDDAS) motivate the development of optimization approaches capable of adapting to streaming, heterogeneous, and asynchronous data from sensor networks. Many established optimization solvers—such as branch-and-bound, gradient descent, and Newton–Raphson methods—rely on iterative algorithms whose step-by-step convergence makes them too slow for real-time, multi-sensor environments. In our recent work, we introduced \LOOPPE (Learning to Optimize the Optimization Process – Permutation Equivariance version), a feed-forward neural approximation model with an integrated feasibility recovery function. \LOOPPE~ processes inputs from a variable number of sensors in arbitrary order, making it robust to sensor dropout, communication delays, and system scaling. Its permutation-equivariant architecture ensures that reordering the input data reorders the corresponding dispatch decisions consistently, without retraining or pre-alignment. Feasibility is enforced via a generalized gauge map, guaranteeing that outputs satisfy physical and operational constraints. We illustrate the approach in a DDDAS-inspired case study of a Virtual Power Plant (VPP) managing multiple distributed generation agents (DERs) to maximize renewable utilization while respecting system limits. Results show that \LOOPPE~ produces near-optimal, feasible, and highly adaptable decisions under dynamic, unordered, and distributed sensing conditions, significantly outperforming iterative-algorithm-based solvers in both speed and flexibility. Here, we extend our earlier work by providing additional analysis and explanation of \LOOPPE’s design and operation, with particular emphasis on its feasibility guarantee and permutation equivariance feature.

\end{abstract}

\section{Introduction}

The rapid growth of sensor-based networks  is a key technological direction that can benefit and also be exploited by Dynamic Data-Driven Applications Systems (DDDAS) \cite{ref_article_0}, a paradigm where real-time data can be continually incorporated into executing system-cognizant computational models to enhance decision-making, prediction, and system optimization. DDDAS-based modeling and instrumentation approaches (or InfoSymbiotic Systems - another term used to denote DDDAS methods and environments), emphasize the synergy between information acquisition and processing, ensuring adaptive responses to changing environments. The DDDAS framework is particularly practical for large-scale, complex systems such as smart power grids, communications grids, transportation networks, autonomous vehicles, and industrial automation, where decisions must be efficient, real-time, and robust to uncertainties.

At the heart of dynamic sensor-based system management lies the frequent need to solve large-scale economic optimization problems in real time \cite{ref_article_1, ref_article_1_1}.  Many established optimization solvers—such
as branch-and-bound, gradient descent, and Newton–Raphson methods—rely on
iterative algorithms whose step-by-step convergence makes them slow for
real-time, multi-sensor environments. As sensor data collection is not continuous and uniform but manifests dynamicity in frequency and modalities of collection, especially when multiple types of sensors are involved, conventional optimization solvers (e.g., GUROBI \cite{gurobi}), require re-optimization, significantly increasing computational overhead.

To address optimization complexity, machine learning (ML)-based optimization is recently being considered as a promising alternative, leveraging historical data to learn direct mappings from input conditions to near-optimal solutions, reducing the number of required iterations \cite{ref_article_2, ref_article_3, ref_article_4}. However, real-world sensor-based decision systems introduce significant challenges that complicate this process. Factors such as sensor dropout due to hard-ware failures, asynchronous communication delays, and the frequent addition or removal of sensors in a dynamic network pose severe difficulties \cite{ref_article_5}. Most conventional ML-based optimization models assume fixed input dimensions, making them incapable of handling dynamically changing sensor networks. In contrast, real-world applications involve models that are order-independent, robust to sensor failures, and adaptable to varying numbers of active sensors, a need that standard deep learning architectures fail to address.

A key challenge in designing ML-based optimizers for DDDAS is ensuring that the model remains robust to variations in sensor availability and ordering. This is where permutation equivariant optimization \cite{ref_pe} plays a crucial role. In sensor-based decision systems, the ordering of input features should not affect the outcome—whether sensor A reports before sensor B should not change the optimization decision. However, standard neural networks inherently assume ordered inputs, leading to inconsistencies when sensors are added, removed, or reordered \cite{ref_article_5.1, ref_article_5.2}.

Permutation equivariance (PE) addresses this by ensuring that if the input sensors are permuted, the output remains correspondingly permuted, preserving the relational structure of the data while remaining invariant to order changes. The PE property enables the model to handle unordered, distributed, and asynchronous data flows, making it highly adaptable for real-time DDDAS operations. Additionally, permutation equivariance ensures that neural approximations can generalize across different sensor configurations, maintaining consistent decision-making even when the network structure changes.

Beyond computational efficiency and adaptability, a critical challenge in applying ML-based optimization to sensor-driven decision systems is ensuring the feasibility of solutions—that is, guaranteeing that every decision satisfies all relevant physical laws, engineering limits, and operational requirements. In practical terms, a feasible solution is one that can be directly implemented in the real system without violating constraints such as generator output limits in power grids, vehicle capacity restrictions in transportation networks, or safety margins in industrial automation. Many existing ML-based optimization approaches struggle to enforce such hard constraints, which are non-negotiable in real-world operations. Traditional techniques for handling constraints, such as penalty methods \cite{ref_article_6} or projection techniques \cite{ref_article_7}, either impose soft constraints—allowing small violations that can still render the solution unusable—or require iterative correction steps, both of which introduce additional computational overhead. As a result, these methods may fail to guarantee strict feasibility while also undermining the speed advantage that motivates ML-based optimization in the first place.

Our prior works \cite{ref_article_8} introduced a non-iterative, feed-forward gauge map approach to ensure that all generated solutions satisfy required constraints without additional optimization steps. The gauge mapping is a differentiable function that performs a radial projection, mapping any arbitrary decision onto the allowable region defined by the constraints. This guarantees that the resulting outputs comply with all specified limits, such as generator capacity bounds or transmission line limits in energy systems. A key advantage of the gauge map method is that it can be applied to any ML regression architecture, providing a flexible mechanism for safe, real-time decision-making in domains where violating operational or physical limits is unacceptable. In the conference paper, we extend the concept from fixed linear constraints to a permutation-equivariant setting, enabling our method to accommodate dynamically changing sensor networks while preserving these constraint-satisfaction guarantees.

To address the aforementioned challenges, we presented the ML-enabled neural approximator named \LOOPPE (Learning to Optimize the Optimization Process—Permutation Equivariance version) in our recent DDDAS2024 conference paper. Our model is specifically designed to:

\begin{itemize}
    \item Dynamically process an arbitrary number of sensor inputs without requiring predefined input dimensions.
    \item Map sensor data to optimal decisions in a permutation equivariant manner, ensuring consistent outputs regardless of sensor order.
    \item Integrate feasibility constraints via the generalized gauge map method, guaranteeing that solutions adhere to physical and engineering limitations.
\end{itemize}

 In this chapter, we extend our earlier work by providing additional analysis and explanation of \LOOPPE’s design and operation, with particular emphasis on its feasibility guarantee and permutation equivariance feature.

\section{Problem Formulation} 

We consider a dynamic sensor network represented by a set of sensors $\mathcal{N}_{\texttt{A}}$, where each sensor is indexed by $i \in \mathcal{N}_{\texttt{A}}$. The roles and functionalities of these sensors evolve over time due to internal system reconfigurations and external environmental variations. Sensors may drop out, be added, or report data in varying orders due to communication delays. To support robust operation under such conditions, we aim to design an optimization framework that can adapt seamlessly to these dynamic changes.

\subsection{Optimization Problem} 

Let $\mathbf{x}^i$ denote the observed input parameters from sensor $i$ and $\mathbf{u}^i$ denote its decision variable. The global optimization objective aggregates individual sensor-level objectives $f^i(\mathbf{u}^i,\mathbf{x}^i)$ as follows:

\begin{subequations}
    \label{distributed compact}
\begin{gather}
    \min  
    f(\mathbf{u},\mathbf{x})=\sum_{i\in \mathcal{N}_{\texttt{A}}}f^i(\mathbf{u}^i,\mathbf{x}^i) \label{distributed compact objective}\\
     \mathbf{u}=\left [ \mathbf{u}^1,...,\mathbf{u}^i,..., \forall i\in \mathcal{N}_{\texttt{A}}  \right ], \quad \mathbf{x}=\left [ \mathbf{x}^1,...,\mathbf{x}^i,..., \forall i\in \mathcal{N}_{\texttt{A}} \right ]  \\
\textup{local constraints: }\left\{\begin{matrix}
\mathbf{A}_{\texttt{eq}}(\mathbf{x}^i)\mathbf{u}^i+\mathbf{B}_{\texttt{eq}}(\mathbf{x}^i)=\mathbf{0}\\ 
\mathbf{A}_{\texttt{ineq}}(\mathbf{x}^i)\mathbf{u}^i+\mathbf{B}_{\texttt{ineq}}(\mathbf{x}^i)\leq\mathbf{0}
\end{matrix}\right. , \forall i\in \mathcal{N}_{\texttt{A}} \label{distributed compact local}\\
\textup{coupled constraints: } \sum_{i\in \mathcal{N}_{\texttt{A}}}[\mathbf{A}(\mathbf{x}^i)\mathbf{u}^i+\mathbf{B}(\mathbf{x}^i)]\leq \mathbf{0}.\label{distributed compact connection}
\end{gather}
\end{subequations}

This formulation captures both local feasibility requirements and system-wide coordination. The local constraints in \eqref{distributed compact local} ensure that each sensor respects its individual operational limits, while the coupled constraints in \eqref{distributed compact connection} enforce coordination among all sensors. We applied variable elimination \cite{ref_article_3} to the equality equations in \eqref{distributed compact local} and then
the sets of constraints in \eqref{distributed compact local} and \eqref{distributed compact connection} can be expressed more generally in a reformulated form: 
\begin{align}
    \sum_{i\in \mathcal{N}_{\texttt{A}}} \mathbf{H}(\mathbf{x}^i)\mathbf{u}^i \leq \sum_{i\in \mathcal{N}_{\texttt{A}}} \mathbf{h}(\mathbf{x}^i). 
\end{align}
Here, $\mathbf{H}(\mathbf{x}^i)$ collects the appropriate rows, while $\mathbf{h}(\mathbf{x}^i)$ stacks the corresponding right-hand side terms. 

This compact form highlights the structural symmetry of the problem: every sensor contributes through the same template $(\mathbf{H}(\cdot),\mathbf{h}(\cdot))$, and the constraints depend only on summations over $i \in \mathcal{N}_{\texttt{A}}$. Consequently, the optimization problem is invariant to any permutation of the sensor indices.

\subsection{Permutation Equivariance}

The invariance described above can be formalized through the concept of \textit{permutation equivariance}. In practice, sensor inputs may be reordered, dropped, or extended due to dynamic conditions. An optimization model must remain consistent under these variations.

Formally, a function $\xi$ is \textit{permutation equivariant} if, for any permutation $\sigma$ over $\mathcal{N}_{\texttt{A}}$:
\begin{align}
    \left [ \mathbf{u}^{\sigma(1)}, \mathbf{u}^{\sigma(2)},...,\mathbf{u}^{\sigma(i)},... \right ]=\xi\left ( \left [ \mathbf{x}^{\sigma(1)}, \mathbf{x}^{\sigma(2)},...,\mathbf{x}^{\sigma(i)},... \right ] \right ).\label{permutation}
\end{align}

This property ensures that the optimizer is agnostic to input ordering while maintaining consistent mappings between each sensor’s input and its decision.

\paragraph{Illustrative Examples.}

Consider a sensor-based power management system where a set of $|\mathcal{N}_{\texttt{A}}|$ distributed energy resources (DERs) provide real-time generation data to a Virtual Power Plant (VPP). Each DER $i$ provides input information $\mathbf{x}^i = (P_{\texttt{C}}^i, P_{\texttt{D}}^i)$, where $P_{\texttt{C}}^i$ represents its generation capacity and $P_{\texttt{D}}^i$ its demand.

\textbf{1) Unfixed Input Order (Delays).} Suppose the system initially receives the data from three DERs in the order:
\begin{align}
    \mathbf{x}^1 = (P_{\texttt{C}}^1, P_{\texttt{D}}^1), \quad \mathbf{x}^2 = (P_{\texttt{C}}^2, P_{\texttt{D}}^2), \quad \mathbf{x}^3 = (P_{\texttt{C}}^3, P_{\texttt{D}}^3),
\end{align}
and produces the dispatch decisions:
\begin{align}
    \mathbf{u}^1,\mathbf{u}^2,\mathbf{u}^3 = \xi(\mathbf{x}^1,\mathbf{x}^2,\mathbf{x}^3).
\end{align}
If data arrives in a different order due to network delays, e.g.,
\begin{align}
    \mathbf{x}^3, \mathbf{x}^1, \mathbf{x}^2,
\end{align}
a permutation-equivariant optimizer must still produce the same decisions but reordered accordingly:
\begin{align}
    \mathbf{u}^3,\mathbf{u}^1,\mathbf{u}^2 = \xi(\mathbf{x}^3,\mathbf{x}^1,\mathbf{x}^2).
\end{align}  

    \textbf{2) Sensor Dropout.} In real-world deployments, a DER may temporarily fail to report its data due to sensor malfunction or communication loss. If $\mathbf{x}^2$ is unavailable, the optimizer instead receives:
\begin{align}
    \mathbf{x}^1, \mathbf{x}^3.
\end{align}
A permutation-equivariant model can still generate correct dispatches for the remaining DERs, $\mathbf{u}^1, \mathbf{u}^3$, without requiring retraining or reconfiguration.
  
\textbf{3) System Scaling.} As the system expands, new DERs may be added. Suppose a fourth DER, $\mathbf{x}^4 = (P_{\texttt{C}}^4, P_{\texttt{D}}^4)$, is introduced:
\begin{align}
    \mathbf{x}^1, \mathbf{x}^2, \mathbf{x}^3, \mathbf{x}^4.
\end{align}
A permutation-equivariant optimizer seamlessly incorporates the new input, producing the corresponding decisions $\mathbf{u}^1, \mathbf{u}^2, \mathbf{u}^3, \mathbf{u}^4$ while maintaining consistency with the established dispatch logic.

\begin{figure}[h]
\centering
\includegraphics[width=0.8\textwidth]{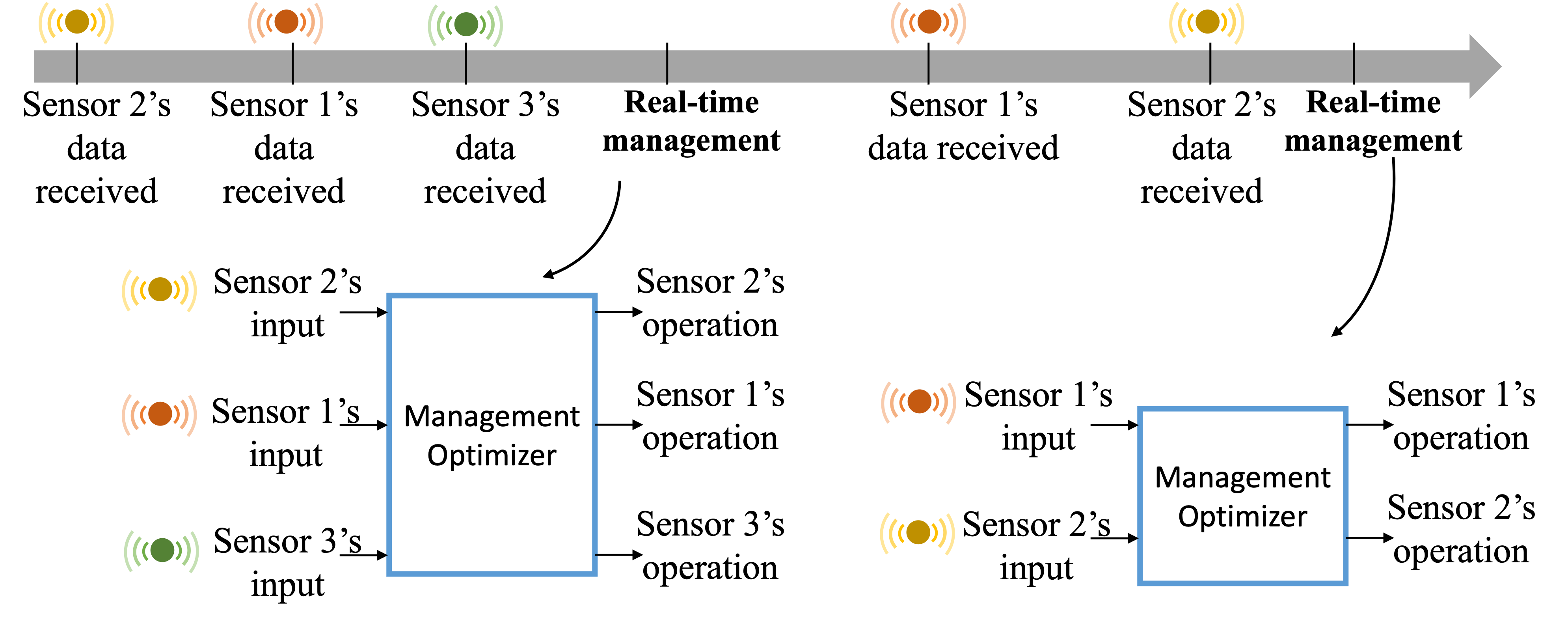}
\caption{Illustration of the permutation equivariance property in the real-time management optimizer. This property guarantees outputs follow the input order, enabling adaptability to sensor addition, dropout, or reordering.}
\label{fig:permutation}
\end{figure}

\subsection{Neural Approximation}

Directly solving the optimization problem using traditional solvers using iterative algorithms can be computationally prohibitive for real-time applications, especially as the number of sensors $|\mathcal{N}_{\texttt{A}}|$ grows or when system conditions change rapidly. To address this challenge, we introduce a neural approximation function $\xi$, which learns to map sensor inputs $\mathbf{x}$ to near-optimal decision variables $\mathbf{u}$:

\begin{align}
    \mathbf{u} = \xi(\mathbf{x}; \Theta),
\end{align}

where $\Theta$ denotes the learnable parameters of the neural model. The goal is for $\xi$ to approximate the optimizer of \eqref{distributed compact} while maintaining essential structural properties, namely \textit{permutation equivariance} and \textit{feasibility preservation}.

\paragraph{Permutation Equivariance in Neural Networks.}
To ensure robustness against varying sensor orderings, the neural function $\xi$ must be permutation equivariant, as formalized in \eqref{permutation}. Concretely, if $\mathbf{x}$ is permuted by an operator $\sigma$, the output $\mathbf{u}$ must be permuted accordingly:

\begin{align}
    \xi(\sigma(\mathbf{x}); \Theta) = \sigma(\xi(\mathbf{x}; \Theta)). \label{nn_equivariance}
\end{align}

\paragraph{Feasibility Preservation.}
In addition to equivariance, it is essential that the neural approximation respects the constraints of the original optimization problem. Specifically, the outputs $\mathbf{u}^i$ must satisfy the local constraints in \eqref{distributed compact local} and the coupled constraints in \eqref{distributed compact connection}. 

By combining permutation-equivariant architectures with feasibility-preserving modules, the neural approximation $\xi$ inherits the essential structural properties of the original optimization formulation, allowing the neural model to generalize across dynamic sensor configurations—handling reordering, dropout,  or scaling—while producing valid and reliable decisions in real time.

\subsection{Connection to the DDDAS paradigm.} 
The above formulation is well-aligned with the DDDAS paradigm. In DDDAS, models and data streams interact in a closed-loop manner: streaming data dynamically reconfigures the model, while model outputs feed back into the physical system to guide operation. Our formulation captures this duality by explicitly accounting for dynamic sensor networks, where data sources may be added, dropped, or reordered in real time. The optimization framework adapts seamlessly to these variations through its permutation-invariant structure, while the decision variables $\mathbf{u}$ directly inform actuation (e.g., dispatch in energy networks). This bidirectional coupling between evolving data and adaptive optimization embodies the core concept of DDDAS.

\section{Proposed Method}

\begin{figure}[h]
\centering
\includegraphics[width=0.9\textwidth]{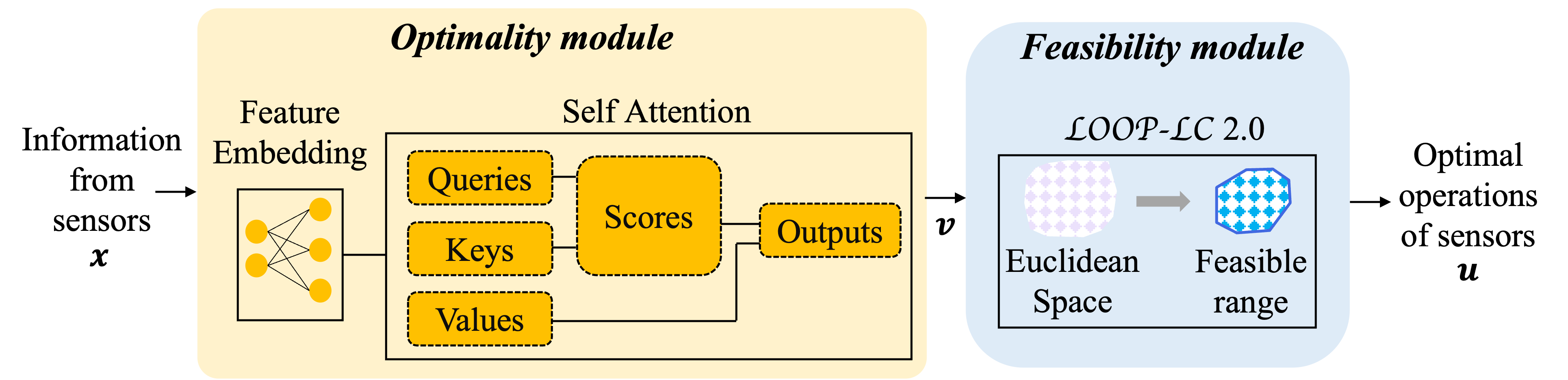}
\caption{Building blocks of the proposed \LOOPPE~ model. The Optimality Module uses an attention mechanism to process input from varying sensor numbers and generate virtual predictions. The Feasibility Module uses the gauge map \cite{ref_article_8} to convert these predictions into practical, constraint-compliant actions, ensuring flexibility and robustness across different sensor setups and dynamics.}
\label{block}
\end{figure}

In this section, we introduce the proposed permutation-equivariant neural approximator \LOOPPE~,  designed to tackle the optimization problem detailed in \eqref{distributed compact}, specifically within a dynamically changing sensor network. Figure \ref{block} illustrates the core components of our method, which include the Optimality and Feasibility Modules.

\subsection{Optimality Module}

The optimality module consumes the data from sensors ($\mathbf{x}$) and output virtual predictions for each($\mathbf{v}$), it consists of two submodules: 

(i) Feature Embedding: Use a fully connected neural network to project the input features of each sensor to a higher dimensional feature space. This is to capture more complex interactions in the subsequent attention mechanism.

(ii) Self Attention: Allow the network to consider and appropriately weigh the importance of each sensor's features relative to others. This self-attention mechanism is particularly beneficial for data where the interaction between sensors plays a more crucial role than their absolute positions in the input sequence.

\subsection{Feasibility module}

The feasibility module utilizes our previous gauge map method \cite{ref_article_8} to convert the predictions $\mathbf{v}$ into practical actions $\mathbf{u}$ that adhere to both local and coupled constraints, maintaining the permutation-equivariance property throughout. 

The gauge map model is capable of mapping the Optimality module's solution to a feasible point within the linearly-constrained
domain.  For equality constraints in \eqref{distributed compact local}, it applies variable elimination to reduce the size of the problem to a reformulated problem only with independent variables. For inequality constraints in \eqref{distributed compact local} and \eqref{distributed compact connection}, a generalized gauge map is adopted to rescale any infeasible solutions to the boundary of the constraint set. The generalized gauge map will keep
the virtual predictions as they are if they have already fallen
within the desired feasible range. Otherwise, it will rescale the virtual predictions to the
boundary. The feasibility module, whose mapping is denoted by $\mathbb{T}$ has the form below:

\begin{align}
    \mathbf{u} = \mathbb{T}(\mathbf{v}) 
    = \mathbf{u}_{\texttt{0}}(\mathbf{x}) +
    \frac{1}{\underset{r}{\max}\left\{\, 1,\; 
    \left[ \frac{\sum_{i\in \mathcal{N}_{\texttt{A}}}\mathbf{H}(\mathbf{x}^i)\mathbf{v}^i}
    {\sum_{i\in \mathcal{N}_{\texttt{A}}}\mathbf{h}(\mathbf{x}^i)} \right]^r 
    \right\}} \mathbf{v} \label{T}
\end{align}

where the summation $\sum_{i \in \mathcal{N}_{\texttt{A}}} \mathbf{H}(\mathbf{x}^i)\mathbf{v}^i$ 
yields a column vector, and 
$\sum_{i \in \mathcal{N}_{\texttt{A}}}\mathbf{h}(\mathbf{x}^i)$ is also a column vector. 
The ratio $\sum_{i \in \mathcal{N}_{\texttt{A}}}\mathbf{H}(\mathbf{x}^i)\mathbf{v}^i/\sum_{i \in \mathcal{N}_{\texttt{A}}}\mathbf{h}(\mathbf{x}^i)$ is computed element-wise, producing a column vector. The notation 
$\max\{1, [\cdot]^r\}$ selects the maximum between 1 and the $r$-th element of this ratio. 

The point $\mathbf{u}_{\texttt{0}}(\mathbf{x})$ is an interior feasible solution, 
which is itself permutation equivariant in $\mathbf{x}$. 
Therefore, $\mathbb{T}$ defines a closed-form, permutation-equivariant feasibility mapping. 
A more detailed derivation is available in \cite{ref_article_8}.

\section{Permutation Equivariance of \LOOPPE}
\paragraph{Setup and notation.}
Let $\mathcal N_{\texttt{A}}=\{1,\dots,n\}$ index the sensors.
Each sensor $i$ provides features $\mathbf x^i\in\mathbb R^{d_x}$, the optimality module outputs a virtual prediction $\mathbf v^i\in\mathbb R^{d_v}$, and the feasibility module returns an action $\mathbf u^i\in\mathbb R^{d_u}$.
We stack per-sensor quantities, e.g.,
$\mathbf X=[\mathbf x^1;\dots;\mathbf x^n]\in\mathbb R^{n\times d_x}$ and similarly $\mathbf V,\mathbf U$.
For any permutation $\sigma\in S_n$, let $P_\sigma\in\{0,1\}^{n\times n}$ denote the corresponding permutation matrix acting on rows, so that
$P_\sigma\mathbf X=[\mathbf x^{\sigma(1)};\dots;\mathbf x^{\sigma(n)}]$, and analogously for $\mathbf V,\mathbf U$.
A mapping $\mathcal F$ is \emph{permutation–equivariant} if
\begin{gather}
   \mathcal F(P_\sigma\mathbf X)=P_\sigma\,\mathcal F(\mathbf X)\quad\text{for all }\sigma\in S_n. 
\end{gather}

The \LOOPPE~model has two components:  
(i) the \emph{optimality module} $\mathcal O:\mathbf X\mapsto\mathbf V$, which consists of a pointwise feature embedding and a self–attention layer without positional encodings; and  
(ii) the \emph{feasibility module} $\mathbb T:\mathbf V\mapsto\mathbf U$ given in \eqref{T}, which enforces local and coupled constraints via a gauge-based radial scaling.  
We prove that both modules are permutation–equivariant, and thus their composition is permutation–equivariant.

\subsection{Optimality module is permutation–equivariant}

\paragraph{(a) Pointwise feature embedding.}
Let $f:\mathbb R^{d_x}\to\mathbb R^{d_e}$ be a fully connected network applied identically to each sensor:
$\mathbf E=[f(\mathbf x^1);\dots;f(\mathbf x^n)]$.
For any $\sigma\in S_n$,
\begin{gather}
[f(\mathbf x^{\sigma(1)});\dots;f(\mathbf x^{\sigma(n)})]
= P_\sigma [f(\mathbf x^1);\dots;f(\mathbf x^n)]
= P_\sigma\,\mathbf E,
\end{gather}
so the embedding is permutation–equivariant.

\paragraph{(b) Self–attention without positional encodings.}
With shared weights $W_Q,W_K,W_V$, define
\begin{gather}
Q=\mathbf E W_Q,\quad K=\mathbf E W_K,\quad V_{\text{value}}=\mathbf E W_V,\\
A=\operatorname{softmax}\!\Big(\tfrac{QK^\top}{\sqrt{d_k}}\Big)\in\mathbb R^{n\times n},\quad
\mathbf V_{\text{att}}=A V_{\text{value}}.
\end{gather}
Under $\mathbf E\mapsto P_\sigma\mathbf E$, we have
$Q\mapsto P_\sigma Q$, $K\mapsto P_\sigma K$, and $V_{\text{value}}\mapsto P_\sigma V_{\text{value}}$.  
Hence
\[
\frac{QK^\top}{\sqrt{d_k}}
\mapsto
P_\sigma\!\Big(\tfrac{QK^\top}{\sqrt{d_k}}\Big)P_\sigma^\top,
\]
and the rowwise softmax satisfies
$\operatorname{softmax}(P_\sigma ZP_\sigma^\top)=P_\sigma\operatorname{softmax}(Z)P_\sigma^\top$.  
Thus $A\mapsto P_\sigma A P_\sigma^\top$ and
\[
\mathbf V_{\text{att}} \mapsto (P_\sigma A P_\sigma^\top)(P_\sigma V_{\text{value}})
= P_\sigma (A V_{\text{value}})
= P_\sigma\,\mathbf V_{\text{att}}.
\]
Further pointwise layers preserve this property, so the full optimality module satisfies
\[
\boxed{\;\mathcal O(P_\sigma\mathbf X)=P_\sigma\,\mathcal O(\mathbf X).\;}
\]

\subsection{Feasibility module is permutation–equivariant}

The feasibility map is
\begin{gather}
\mathbb T(\mathbf V)=\mathbf u_{\texttt 0}(\mathbf X)+c(\mathbf X,\mathbf V)\mathbf V,
\end{gather}
with scaling factor
\begin{gather}
c(\mathbf X,\mathbf V)=
\Bigg[\max_{r}\Bigg\{1,\Bigg[\frac{\sum_{i=1}^n\mathbf H(\mathbf x^i)\mathbf v^i}
{\sum_{i=1}^n\mathbf h(\mathbf x^i)}\Bigg]^r\Bigg\}\Bigg]^{-1}.
\end{gather}

\paragraph{(a) Permutation–invariance of the scaling factor.}
For any $\sigma$, reordering $(\mathbf x^i,\mathbf v^i)$ as $(\mathbf x^{\sigma(i)},\mathbf v^{\sigma(i)})$ does not affect the sums:
\[
\sum_{i=1}^n\mathbf H(\mathbf x^{\sigma(i)})\mathbf v^{\sigma(i)}
=\sum_{i=1}^n\mathbf H(\mathbf x^i)\mathbf v^i,\qquad
\sum_{i=1}^n\mathbf h(\mathbf x^{\sigma(i)})=\sum_{i=1}^n\mathbf h(\mathbf x^i).
\]
Thus $c(P_\sigma\mathbf X,P_\sigma\mathbf V)=c(\mathbf X,\mathbf V)$.

\paragraph{(b) Equivariance of $\mathbb T$.}
Since $\mathbf u_{\texttt 0}(\mathbf X)$ is permutation–equivariant by assumption,
\begin{align}
\mathbb T(P_\sigma\mathbf V)
&=\mathbf u_{\texttt 0}(P_\sigma\mathbf X)+c(P_\sigma\mathbf X,P_\sigma\mathbf V)(P_\sigma\mathbf V) \\
&=P_\sigma\mathbf u_{\texttt 0}(\mathbf X)+c(\mathbf X,\mathbf V)P_\sigma\mathbf V \\
&=P_\sigma\!\bigl[\mathbf u_{\texttt 0}(\mathbf X)+c(\mathbf X,\mathbf V)\mathbf V\bigr] \\
&=P_\sigma\,\mathbb T(\mathbf V).
\end{align}
Hence the feasibility module is permutation–equivariant.

\subsection{End-to-end permutation–equivariance}

Let $\mathcal F=\mathbb T\circ\mathcal O$ denote the full \LOOPPE~mapping.
For any $\sigma\in S_n$,
\begin{align}
\mathcal F(P_\sigma\mathbf X)
=\mathbb T(\mathcal O(P_\sigma\mathbf X))
=\mathbb T(P_\sigma\mathcal O(\mathbf X))
=P_\sigma\mathbb T(\mathcal O(\mathbf X))
=P_\sigma\,\mathcal F(\mathbf X).
\end{align}
Thus \LOOPPE~is permutation–equivariant.

\begin{theorem}[Permutation–equivariance of \LOOPPE]
\label{thm:perm-equivar}
Assume:  
(i) the feature embedding applies the same network to each sensor independently;  
(ii) self–attention uses shared $W_Q,W_K,W_V$ and no positional encodings;  
(iii) the interior point $\mathbf u_{\texttt 0}(\mathbf X)$ is permutation–equivariant in $\mathbf X$; and  
(iv) the feasibility map $\mathbb T$ has the gauge form \eqref{T}.  
Then the full mapping $\mathcal F=\mathbb T\circ\mathcal O$ is permutation–equivariant:
\[
\forall\sigma\in S_n,\qquad
\mathcal F(P_\sigma\mathbf X)=P_\sigma\,\mathcal F(\mathbf X).
\]
\end{theorem}

\section{ Experiment results}

\subsection{Experiment Setup}

\subsubsection{Problem:}

To evaluate the effectiveness of the proposed method, we implemented it in a case study involving a Virtual Power Plant (VPP) tasked with managing the assets of 20 distributed generation agents, collectively denoted as $\mathcal{N}_{\texttt{A}}$. These agents, which represent distributed energy resources (DERs), are integrated and coordinated with power grid operations to optimize the utilization of distributed energy while maintaining operational feasibility. The objective of real-time energy management within the VPP is to maximize the utilization of available distributed energy while ensuring that generation limits and system-wide power constraints are met.

The optimization problem governing the VPP operation is formulated as follows:

\begin{subequations}
    \label{vpp}
    \begin{gather}
        \min \sum_{i \in \mathcal{N}_{\texttt{A}}} \left( P_{\texttt{G}}^i - P_{\texttt{C}}^i \right)^2
        \label{vpp_objective} \\
        0 \leq P_{\texttt{G}}^i \leq P_{\texttt{C}}^i, \forall i \in \mathcal{N}_{\texttt{A}}
        \label{vpp_generation} \\
        -P_{\texttt{omax}} \leq \sum_{i \in \mathcal{N}_{\texttt{A}}} (P_{\texttt{G}}^i - P_{\texttt{D}}^i) \leq P_{\texttt{omax}}
        \label{vpp_system}
    \end{gather}
\end{subequations}

where $P_{\texttt{G}}^i$ represents the power generation of agent $i$, $P_{\texttt{C}}^i$ denotes its generation capacity, and $P_{\texttt{D}}^i$ represents the load demand. The objective function in \eqref{vpp_objective} minimizes the deviation between the generated power and the available capacity, thereby penalizing any underutilization of resources. The constraint in \eqref{vpp_generation} ensures that generation remains within the capacity limits of each agent, while the system-wide constraint in \eqref{vpp_system} regulates the net output of the VPP, ensuring that the total power imbalance remains within a predefined operational threshold $P_{\texttt{omax}}$. 

This optimization formulation is inherently aligned with the principles of permutation equivariance. Each agent's contribution to the objective function and constraints is independent of its indexing, ensuring that the problem structure remains valid under any reordering of sensors. This property is essential for handling dynamic environments where the number of active agents may vary due to network conditions, sensor failures, or the addition of new resources.

\subsubsection{Data Generation and Parameter Selection:}

To ensure a realistic evaluation, the agent parameters are selected based on values reported in \cite{ref_article1}, where each agent's generation capacity $P_{\texttt{C}}^i$ varies between $10$ kW and $25$ kW. The maximum allowable power deviation, $P_{\texttt{omax}}$, is set to $100$ kW, imposing operational limits on the aggregated output of the VPP. To introduce variability and assess the robustness of the proposed method, a $10\%$ fluctuation is applied to each agent’s generation capacity and demand.

To simulate realistic real-time operational conditions, we generate a dataset consisting of $400$ samples, with each sample representing a unique combination of generation and demand parameters. Among these, $100$ samples are reserved as test data points for performance evaluation. In each sample, only a random subset of sensors receives the corresponding agent's input parameters, reflecting practical scenarios where sensor measurements may be incomplete or delayed due to communication constraints.

\subsubsection{Benchmarking and Evaluation:}

For performance evaluation, the proposed permutation-equivariant neural optimizer is compared against a conventional optimization approach using the commercial solver GUROBI \cite{gurobi}. GUROBI is employed as the baseline solver due to its widespread use in power system optimization and its ability to provide near-optimal solutions within reasonable computational time. The comparison assesses both computational efficiency and solution quality, focusing on key performance metrics such as execution time, optimality gap, and feasibility adherence.

The evaluation aims to demonstrate the capability of the proposed approach to provide real-time decision-making under dynamically changing conditions. Specifically, the study investigates whether the neural-based optimizer can maintain consistent and feasible power dispatch decisions across different sensor configurations while significantly reducing computational overhead compared to traditional solvers. The subsequent sections present the results and discuss their implications for real-time power system optimization.

\subsection{Results}

The performance of the proposed permutation-equivariant neural optimization method was evaluated in comparison to the conventional solver GUROBI. The evaluation focuses on three key aspects: computational efficiency, optimality gap, and feasibility adherence. The results demonstrate that the proposed method significantly reduces computation time while maintaining solution quality and feasibility under varying sensor configurations.

\subsubsection{Computational Efficiency}

The computational performance of the proposed method is summarized in Table~\ref{tab:time_results}. Due to the dynamic nature of sensor availability, the scale of the optimization problem fluctuates across different instances, impacting the computational time required to reach an optimal solution. Despite these variations, the proposed method consistently achieves significantly lower execution times than the baseline solver. 

\begin{table}[htbp]
\caption{Computational Time Comparison}\centering
\begin{tabular}{|c|c|c|}
\hline
Performance Metric & Gurobi Solver Time (ms) & Proposed Method Time (ms) \\ \hline
Average            & 6.48                     & 0.33                        \\ \hline
Minimum            & 5.02                      & 0.30                      \\ \hline
Maximum            & 24.34                     & 0.58                      \\ \hline
\end{tabular}
\label{tab:time_results}
\end{table}

As observed in Table~\ref{tab:time_results}, the proposed method achieves an average execution time of 0.33 ms, which is approximately 20 times faster than GUROBI. The worst-case execution time of the proposed method remains under 1 ms, whereas GUROBI exhibits significant fluctuations in solution time, reaching a maximum of 24.34 ms. This substantial improvement in computational efficiency underscores the capability of the proposed method to facilitate real-time decision-making, particularly in large-scale sensor-based optimization problems where traditional solvers struggle with high computational overhead.

\subsubsection{Solution Optimality and Feasibility}

To assess the quality of the solutions generated by the proposed method, Table~\ref{tab:gaps} presents the optimality gap and feasibility gap in comparison to the baseline solver. The optimality gap quantifies the deviation of the obtained solutions from the optimal benchmark, defined as:

\begin{align}
    \text{Optimality Gap} = \frac{\left\|\mathbf{u} - \mathbf{u}^*\right\|^2}{\left\|\mathbf{u}^*\right\|^2},
\end{align}

where $\mathbf{u}^*$ represents the optimal solution obtained using GUROBI, and $\mathbf{u}$ denotes the solution generated by the proposed method. The feasibility gap measures constraint violations, ensuring that the generated solutions adhere to both local and coupled constraints.

\begin{table}[htbp]
\caption{Optimality and Feasibility Gaps of the \LOOPPE~ Method.}
\centering
\begin{tabular}{|c|ccc|c|}
\hline
\multirow{2}{*}{Metric} & \multicolumn{3}{c|}{Optimality Gap}                                   & Feasibility Gap \\ \cline{2-5} 
                        & \multicolumn{1}{c|}{Average} & \multicolumn{1}{c|}{Minimum} & Maximum & Minimum         \\ \hline
Compared against baseline                   & \multicolumn{1}{c|}{0.04}    & \multicolumn{1}{c|}{0.00}    & 0.13    & 0.00            \\ \hline
\end{tabular}
\label{tab:gaps}
\end{table}

The results in Table~\ref{tab:gaps} demonstrate that the proposed method maintains a low optimality gap, with an average deviation of 4\% from the optimal solution and a maximum deviation of 13\%. In many instances, the proposed method achieves a 0\% optimality gap, highlighting its capability to approximate optimal solutions effectively.

Furthermore, the feasibility gap remains consistently at 0.00, confirming that the proposed method strictly adheres to operational constraints. Unlike many ML-based optimization approaches that require post-processing to enforce feasibility, the integration of the generalized gauge map ensures that the solutions generated by \LOOPPE~ inherently satisfy the problem constraints. This property is particularly advantageous in real-time applications, where constraint violations can lead to operational inefficiencies or system instability.

\subsubsection{Solution Spectrum Analysis}

\begin{figure}[h]
\centering
\includegraphics[width=0.9\textwidth]{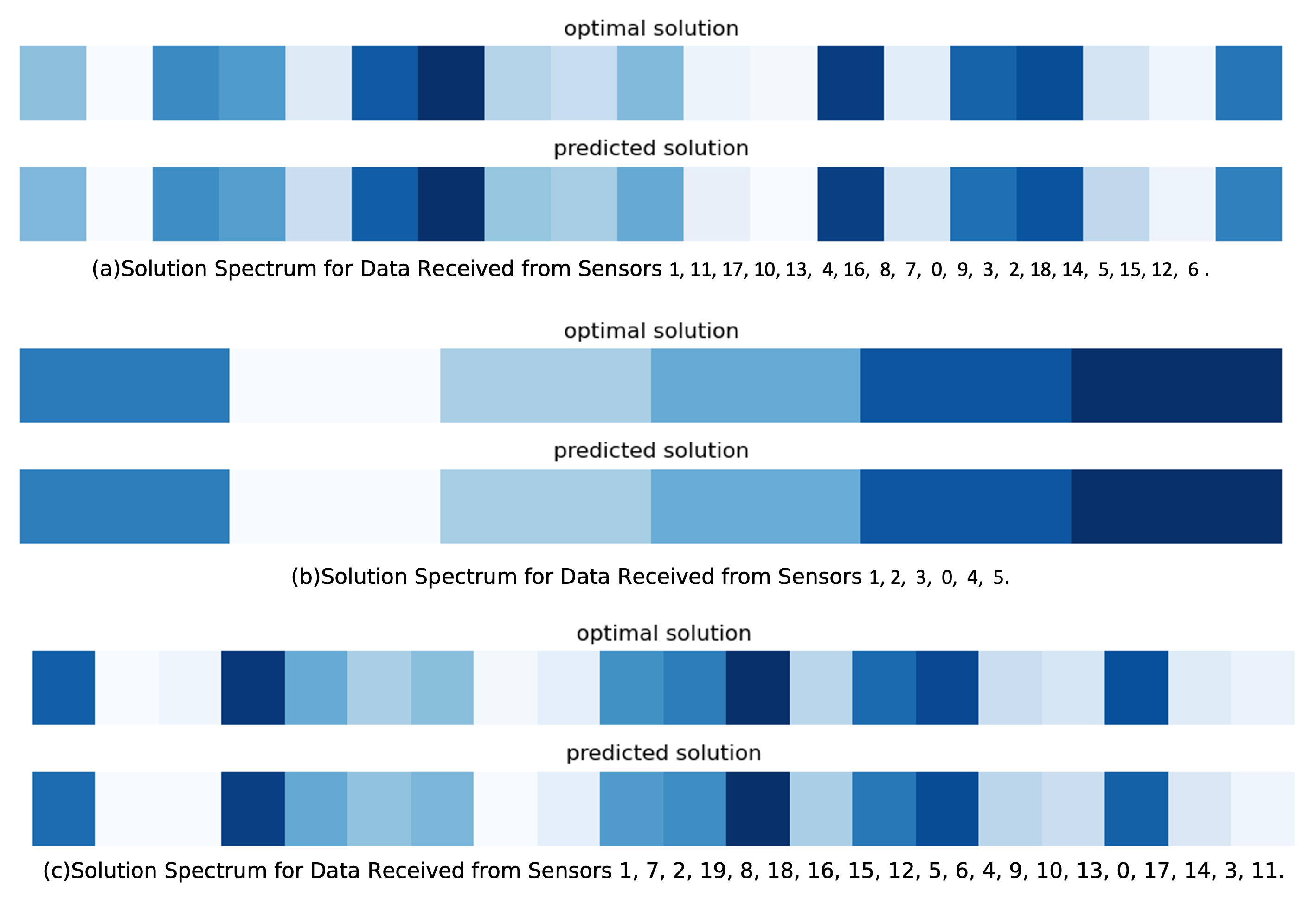}
\caption{Solution spectrum for various test samples using data from multiple sensors. The similarity between the spectra obtained using \LOOPPE~ and the traditional solver highlights the accuracy and reliability of the proposed method. The model's ability to process unordered sensor inputs further enhances its adaptability in dynamic environments.}
\label{fig:spectrum_comparison}
\end{figure}

To further evaluate the effectiveness of the proposed approach, Figure~\ref{fig:spectrum_comparison} presents a comparative analysis of the solution spectra obtained from multiple test samples. The figure illustrates the alignment between the solutions produced by the proposed method and those generated by GUROBI.

As observed in Figure~\ref{fig:spectrum_comparison}, the solution spectra of the proposed method closely resemble those of the baseline solver, demonstrating that it can effectively approximate the optimal solution. The ability of the method to consistently produce similar operational profiles to GUROBI, despite significantly reduced computation times, reinforces its suitability for real-time decision-making in dynamic sensor-based optimization problems.

\subsubsection{Discussion}

The experimental results provide strong empirical evidence of the advantages of using a permutation-equivariant neural optimization approach. The key findings from the evaluation are as follows:

1. The proposed method achieves a significant reduction in computational time, making it well-suited for real-time applications where fast decision-making is required.

2. The low optimality gap indicates that the proposed method provides solutions that are close to the global optimum while maintaining efficiency.

3. The zero feasibility gap confirms that the solutions adhere to all operational constraints without requiring additional constraint-handling mechanisms.

4. The solution spectra comparison demonstrates the strong alignment between the proposed method and traditional solvers, reinforcing its reliability.

Overall, these results highlight the potential of permutation-equivariant neural approximations to replace conventional solvers in dynamic sensor-based decision-making tasks. The ability of the method to generalize across varying sensor configurations while preserving both computational efficiency and solution quality suggests promising applications in domains requiring real-time optimization, such as power systems, autonomous control, and large-scale sensor networks.

\section{Conclusion}

In our conference paper, we introduce a permutation-equivariant neural approximator, \LOOPPE, designed to optimize operations within dynamic and evolving sensor networks. The proposed method addresses the fundamental challenges in real-time optimization by leveraging permutation equivariance to ensure that sensor inputs can be processed in an order-invariant manner. This capability is particularly valuable in Dynamic Data-Driven Applications Systems (DDDAS), where sensor availability may fluctuate due to failures, network delays, or environmental changes. By incorporating permutation-equivariant structures, \LOOPPE~ efficiently adapts to these variations, providing robust, scalable, and computationally efficient capabilities.

The proposed approach builds upon our previous work \cite{ref_article_8} by integrating physical constraints directly within the neural approximation. Traditional ML-based optimizers often struggle with feasibility, requiring post-processing or projection methods to enforce constraints. In contrast, \LOOPPE~ employs a generalized gauge map to ensure that all generated solutions inherently satisfy system constraints, thereby enhancing the practicality of neural approximations in real-world applications. The experimental results validate the effectiveness of our approach, demonstrating that \LOOPPE~ not only achieves near-optimal performance but also significantly reduces computational time compared to conventional solvers such as GUROBI. These findings underscore the potential of neural approximators to replace traditional optimization methods in large-scale, time-sensitive decision-making processes.

Despite these advancements, several challenges and opportunities for future research remain. One key area for exploration is the enhancement of robustness against data uncertainties. Sensor-based decision-making systems are often subject to noise, missing data, or adversarial perturbations, which can degrade optimization performance. Developing uncertainty-aware learning frameworks, such as Bayesian neural networks or robust optimization techniques, could further improve the resilience of permutation-equivariant models. Additionally, extending \LOOPPE~ to decentralized or distributed architectures would enhance its scalability, enabling real-time coordination among multiple decision-making agents in large-scale networks.

Future research should also focus on expanding the model’s applicability to more complex and heterogeneous data types. Many real-world sensor networks involve multimodal data sources, including time-series measurements, categorical metadata, and spatially correlated information. Integrating advanced feature fusion techniques, such as graph neural networks or transformer-based models, could enable \LOOPPE~ to process richer input representations while maintaining its permutation-equivariant properties. Furthermore, applying this framework to emerging fields such as smart grids, autonomous systems, and industrial automation could unlock new avenues for improving efficiency and decision-making in dynamically changing environments. By addressing these research directions, future work can further refine the capabilities of neural approximators and extend their impact across diverse domains.

\section{Acknowledgment}

This research is funded under AFOSR grants \#FA9550-24-1-0099 and FA9550-23-1-0203.

\end{document}